\documentclass[useAMS,usenatbib]{mn2e}
\usepackage{amssymb}
\usepackage{graphicx}
\usepackage{url}
\usepackage[usenames]{color}
\usepackage{url}

\newcommand{\de} {{\rm d}}

\begin{document}

\title[Spectra in OG models]{An assessment of the pulsar outer gap model.\\
II: Implications for the predicted $\gamma$-ray spectra}

\author[D. Vigan\`o et al.]{Daniele Vigan\`o$^1$, Diego F.~Torres$^{1,2}$, Kouichi Hirotani$^3$ \& Mart\'in E.  Pessah$^4$\\ 
$^1$Institute of Space Sciences (CSIC--IEEC), Campus UAB, Faculty of Science, Torre C5-parell, E-08193 Barcelona, Spain\\
$^2$Instituci\'o Catalana de Recerca i Estudis Avan\c{c}ats (ICREA) Barcelona, Spain\\
$^3$Academia Sinica, Institute of Astronomy and Astrophysics (ASIAA), PO Box 23-141, Taipei, Taiwan\\
$^4$Niels Bohr International Academy, Niels Bohr Institute, Blegdamsvej 17, DK-2100, Copenhagen \O, Denmark}

\date{}
\maketitle

\label{firstpage}

\begin{abstract}
One of the most important predictions of any gap model for pulsar magnetospheres is the predicted $\gamma$-ray spectra. In the outer gap model, the properties of the synchro-curvature radiation are sensitive to many parameters, whose realistic ranges have been studied in detail in an accompanying paper. There we demonstrated that the uncertainty in the radius of curvature, the magnetic field geometry, and the X-ray surface flux may affect by orders of magnitude the predicted flux and spectral peak in the $\gamma$-ray regime. Here, we present a systematic, numerical study of the impact of the different parameters on the particle dynamics along the gap and calculate the emitted synchro-curvature radiation along the trajectory. By integrating the emitted radiation along the gap and convolving it with a parametrized particle distribution, we discuss how the comparison with the wealth of {\em Fermi}-LAT data can be used to constrain the applicability of the model. The resulting spectra show very different energy peaks, fluxes and shapes, qualitatively matching the great variety of the observed {\em Fermi}-LAT pulsars. In particular, if we see a large fraction of photons emitted from the initial part of the trajectory, we show that the spectra will be flatter at the low-energy {\it Fermi}-LAT  regime (100 MeV -- 1 GeV). This provides a solution for such observed flat spectra, while still maintain synchro-curvature radiation as the origin of these photons.

\end{abstract}

\section{Introduction}

The $\gamma$-ray detections of pulsars have been steadily increasing during the last few years, reaching more than one hundred sources \citep{2fpc}. This wealth of data allows to quantitatively analyze the different gap models. The light curves, for instance, have been used to constrain the emission sites in the magnetosphere (e.g., \citealt{watters09,pierbattista15}). In this paper, we focus on the contribution of the synchro-curvature radiative losses to the $\gamma$-ray spectra, within the outer gap (OG) model \citep{cheng86a,cheng86b,zhang97}.

In an accompanying paper (\citealt{paper1}, hereafter referred to as Paper I), we have focused on the assumptions underlying the OG model. Here we evaluate the impact of such uncertainties on the predicted $\gamma$-ray spectra. In particular, we study whether the variety of the observed spectra can constrain the model parameters, like the radius of curvature, or the typical Lorentz factor of particles, assuming that the detected radiation originates from synchro-curvature losses. 

The paper is organized as follows. In \S\ref{sec:pair_production} we calculate and discuss the optical depth of the pair production process via photon-photon interaction. In \S\ref{sec:spectrum} we numerically solve the dynamics of a particle in the outer gap, and, using such trajectories, we calculate the $\gamma$-ray spectrum considering the position-dependent values of the different parameters. This is a new approach to the spectral calculations, which usually rely on analytical approximations and does not consider the initial part of the trajectory where synchrotron losses can be important. We also rely on the accompanying study, Paper I, as well as on the synchro-curvature loses discussed by \cite{cheng96}, but expressed under a new formulation, presented in \cite{paper0}. In \S\ref{sec:results} we present and discuss the resulting spectra for different models. In \S\ref{sec:conclusions} we discuss the impact of the results, and the future applications of this study.

%

\begin{figure*}
\centering
\includegraphics[width=.4\textwidth]{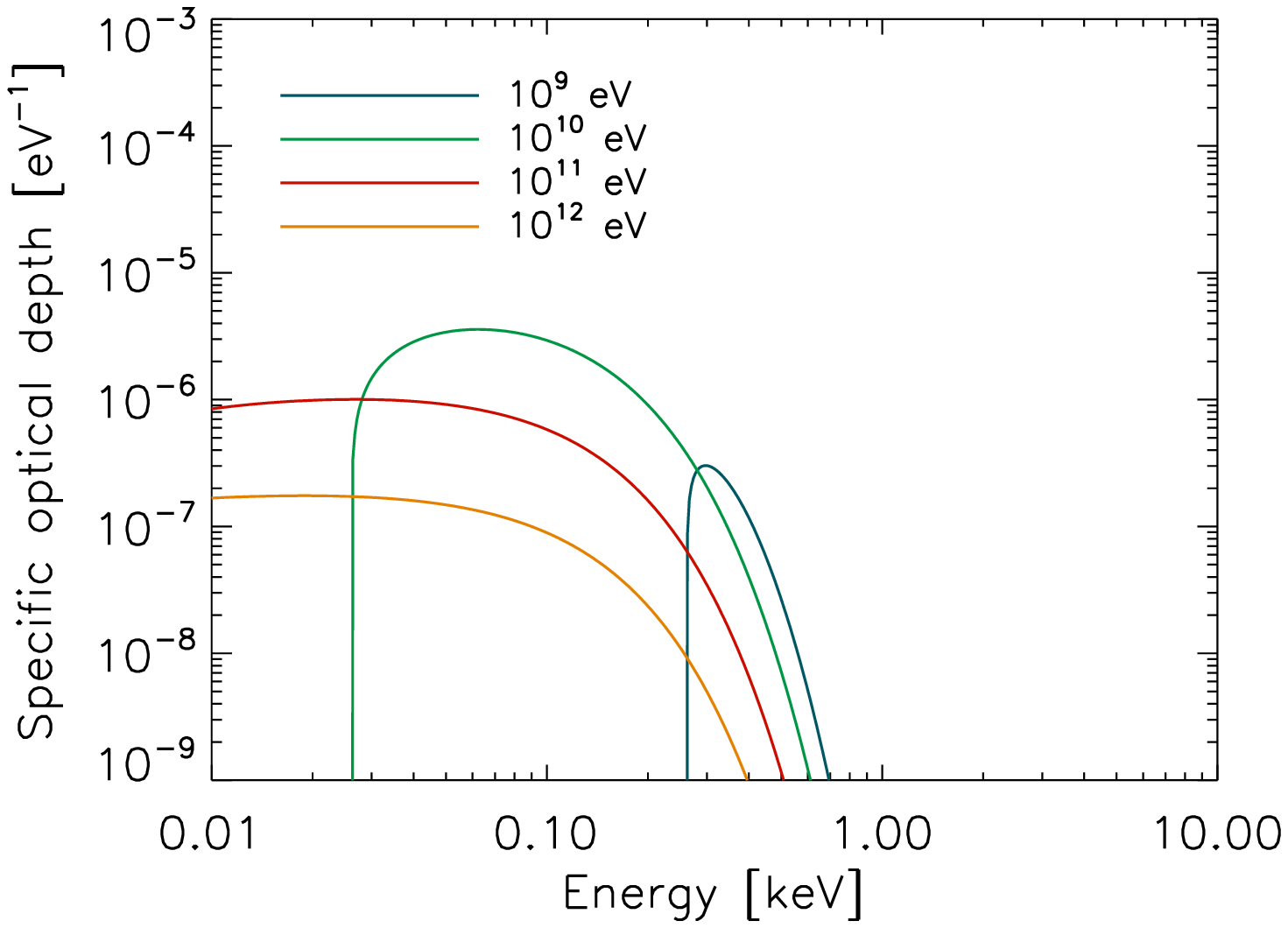}
\includegraphics[width=.4\textwidth]{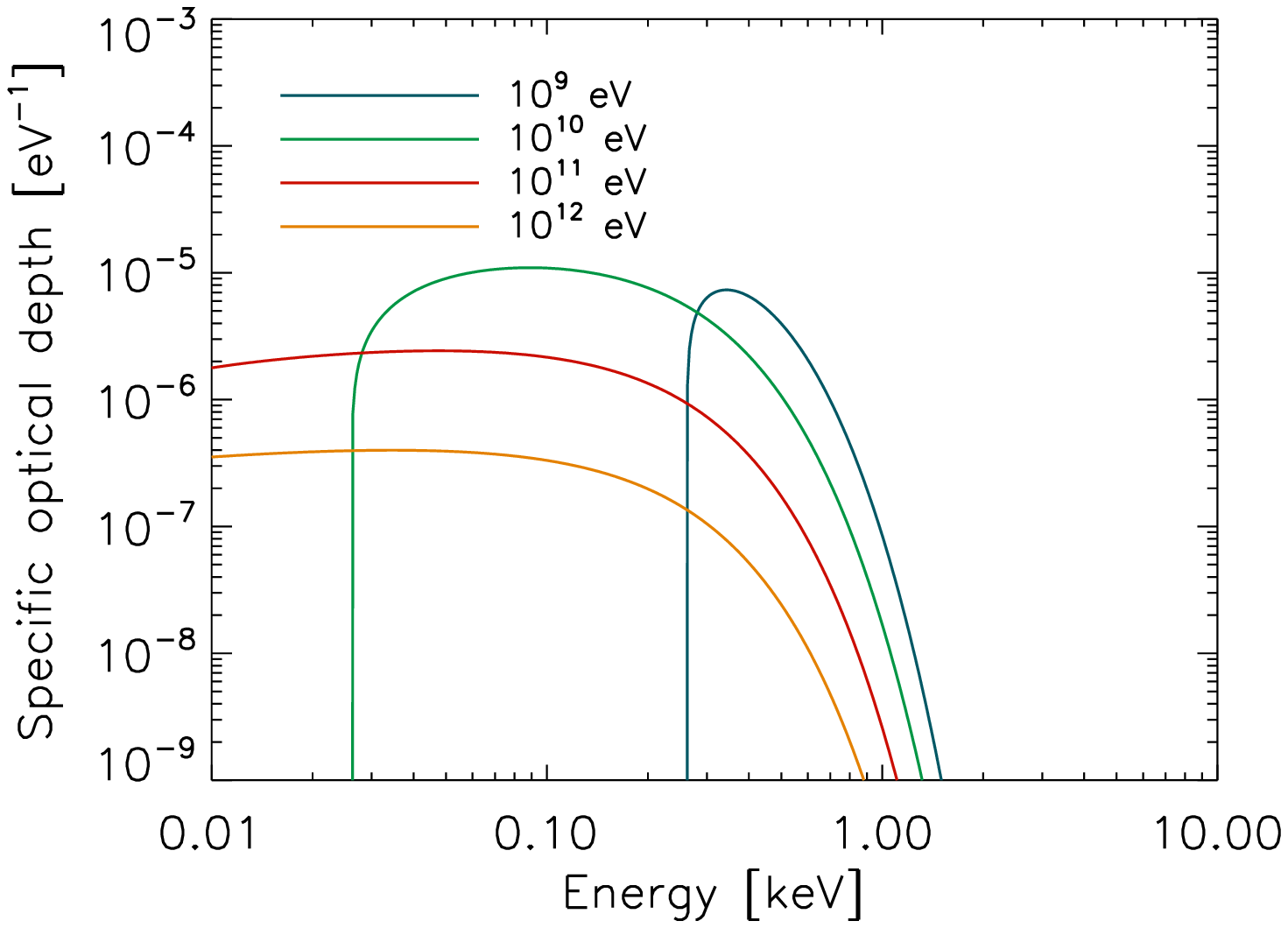}\\
\includegraphics[width=.4\textwidth]{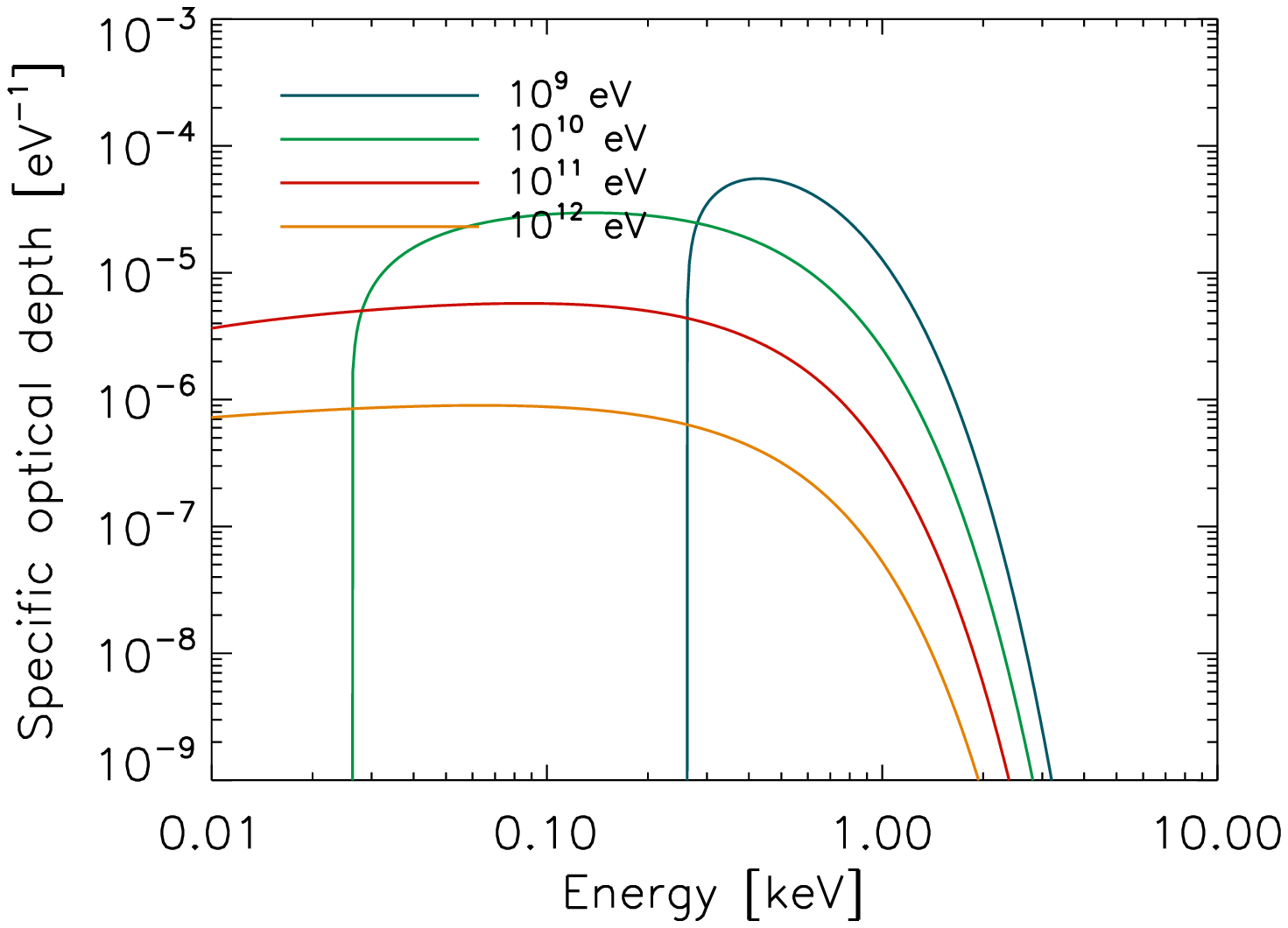}
\includegraphics[width=.4\textwidth]{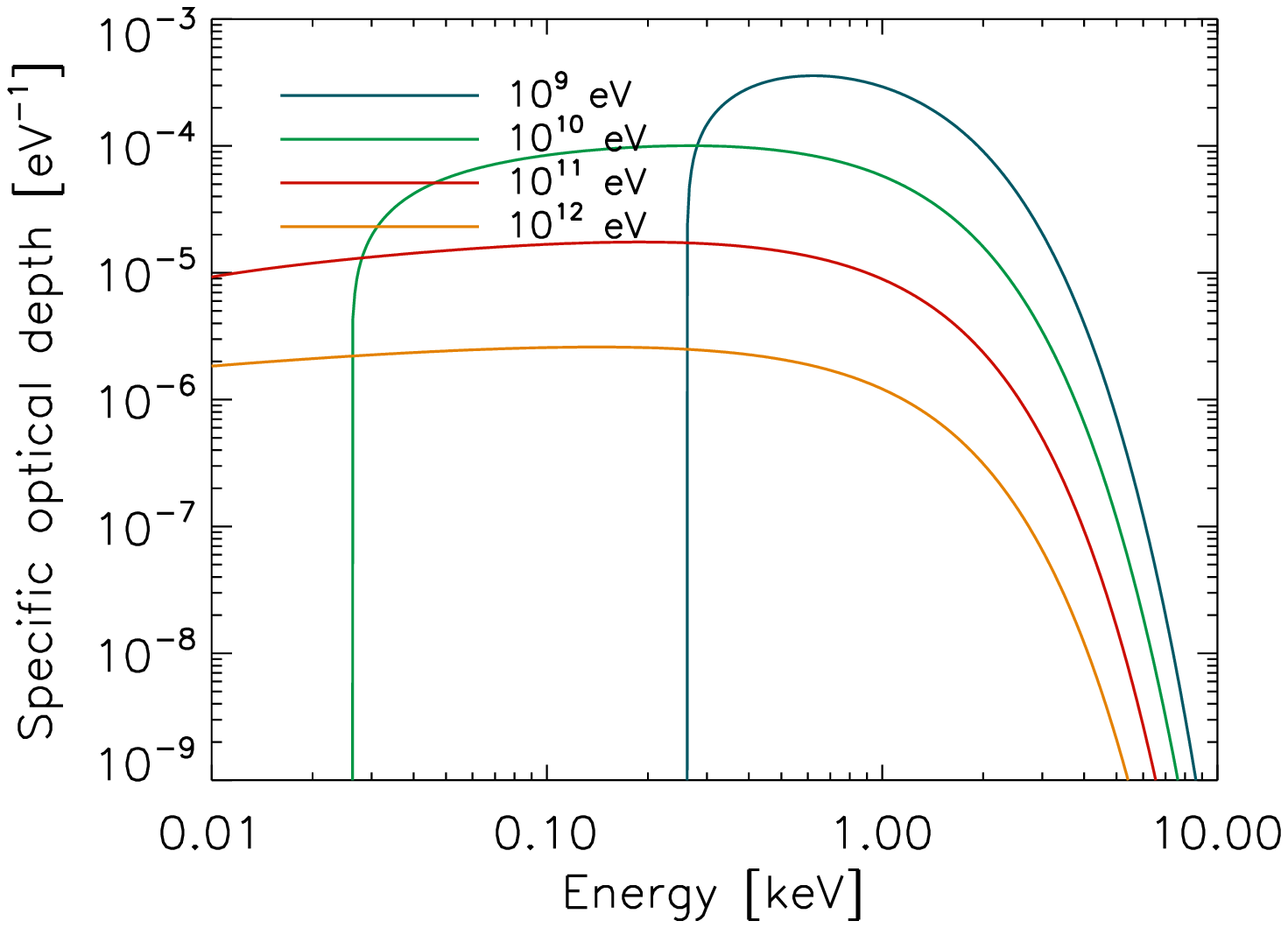}
\caption{Specific optical depth for photon-photon pair production for a $\gamma$-ray photon, Eq.~(\ref{eq:tau_specific}), for different surface temperatures: from left to right $kT=100, 150$ eV (top), 200 and 500 eV (bottom). We assume face-on scattering, i.e., $\cos\psi=-1$, and $r_{\rm in}=2.5\times 10^8$ cm, $r_{\rm out}=5\times 10^8$ cm, and $r_{\rm bb}=10$ km. The different colors indicate the $\gamma$-ray energy, taken in the range $10^9-10^{12}$ eV.}
 \label{fig:tau_spec}
\end{figure*}

\section{Pair production}\label{sec:pair_production}

We consider an OG supported by the interaction between the $\gamma$-ray photons produced in the gap and the soft X-ray surface photons emitted by the hot neutron star (NS) surface. The mechanism sustaining the gap is self-consistent if enough pairs can be produced and emit curvature $\gamma$-ray photons, a fraction of which, in turn, scatter against X-rays. The efficiency of the pair production basically depends on the energies and fluxes of the photons.

The cross section of the $\gamma$-$\gamma$ pair production, for two photons having energies $E_1$, $E_2$ is  \citep{gould67}:
\begin{eqnarray}\label{eq:pp_cross_section}
&& \sigma^{\gamma\gamma}= \frac{3\sigma_T}{16} (1-\mu^2)\times \nonumber\\
  && \hspace{1.3cm} \left[2\mu(\mu^2-2) + (3-\mu^4)\ln\left(\frac{1+\mu}{1-\mu}\right)\right]~,
\end{eqnarray}
where we have used the Thomson cross section
\begin{equation}\label{eq:thomson}
  \sigma_T = \frac{8\pi}{3}\left(\frac{e^2}{mc^2}\right)^2 = 6.65\times10^{-25} \mbox{cm}^2~,
\end{equation}
$e$ is the elementary charge, $m$ the electron mass, and $c$ the speed of light, and

\begin{equation}\label{eq:beta_pair_production}
\mu=\sqrt{1-\frac{2(m c^2)^2}{(1-\cos \psi) E_1 E_2}} ~. 
\end{equation}
Here, $\psi$ is the angle between the propagation directions for the two photons. The maximum value of the cross section is obtained for $\mu_{\rm max}=0.701$ (which means $1-\mu^2_{\rm max}\simeq 0.5$, see also \S2.5 of \citealt{paper1} for further discussion). The scattering is assumed to take place between thermal X-rays proceeding from the surface and $\gamma$-ray photons emitted by the accelerated particles. The optical depth along the gap is
\begin{equation}\label{eq:tau}
 \tau_{\gamma\gamma}(E_\gamma) = \int_{E_X} \int_{l_\gamma} \frac{\de \tau_{\gamma\gamma}}{\de E} ~{\rm d}l_\gamma~{\rm d}E~,
\end{equation}
where the specific optical depth (i.e., the optical depth per unit energy) is
\begin{equation}\label{eq:tau_specific}
 \frac{\de \tau_{\gamma\gamma}}{\de E}(E,E_\gamma) = (1-\cos\psi)\frac{dn_X}{dE}(E,r) \sigma_{\gamma\gamma}(E,E_\gamma)~,
\end{equation}
and the integral in Eq.~(\ref{eq:tau}) is performed along the line traced by $\gamma$-ray photons inside the gap, $l_\gamma$. Assuming that the NS surface emit isotropically as a blackbody with a uniform temperature $T$ over the emitting region $r_{\rm bb}$, the X-ray photon density distribution is
\begin{equation}
 \frac{dn_X}{dE}(E,r)=\left(\frac{r_{\rm bb}}{r}\right)^2\frac{2\pi}{(hc)^3} \frac{E^2}{e^{E/kT}-1}~.
\end{equation}
The temperature can be fixed by considering the long-term cooling, which mostly depends on the age and on many uncertain properties (envelope composition, interior magnetic field configuration, superfluidity, equation of state, etc.). Alternatively, one can use the estimated temperature of the heated polar cap, $T_h$. This value, in turn, contains dependences on $B_\star$, $P$, and, to a lesser extent, on microphysical parameters like the local pair multiplicity, the geometry and the bombardment efficiency. Both of these choices, and the ranges expected for $T$ are discussed in Paper I.

We now calculate the optical depth for a given X-ray flux. For simplicity, we assume a $\gamma$-ray photon propagating inward radially, between the boundaries $r_{\rm in}$ and $r_{\rm out}$, with head-on collisions, $\psi=\pi$. In the following results, we also fix $r_{\rm in}=2.5\times 10^8$ cm, $r_{\rm out}=5\times 10^8$ cm, and $r_{bb}=10$ km. The gap is sustained if, for each primary pair, a large number $N_\gamma$ of $\gamma$-ray photons are created, so that $\tau_{\gamma\gamma} N_\gamma \sim 1$ for a given energy $E_\gamma$. If the pairs are too many, then the gap size will be reduced due to screening. On the other hand, if the pairs are too few, then the gap will grow larger. A precise estimate of $N_\gamma$ must consider the complete cascade dynamics and radiative processes. The simulations by \cite{hirotani13} find that the outflowing photons are more numerous, but less likely to interact (almost tail-on scattering), compared to the inflowing photons.

In each panel of Fig.~\ref{fig:tau_spec}, we show the specific optical depth resulting from Eq.~(\ref{eq:tau_specific}) for a given surface temperature, for different $\gamma$-ray energies, ranging from $10^9$ to $10^{12}$ eV, represented by different color lines. For each of them, the steep vertical drop at low energies represent the energy threshold, for which $\mu=0$ and the cross-section vanishes, Eq.~(\ref{eq:pp_cross_section}).

\begin{figure}
\centering
\includegraphics[width=.45\textwidth]{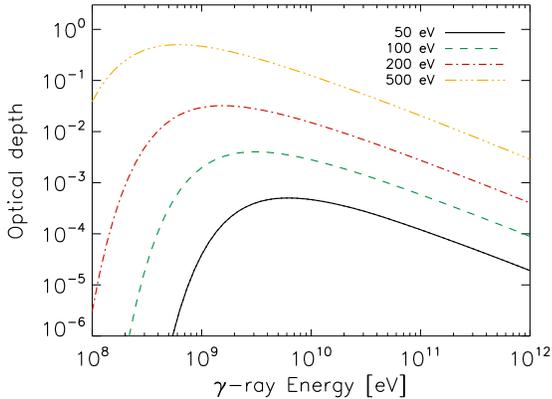}
\caption{Comparison of $\tau_{\gamma\gamma}$, Eq.~(\ref{eq:tau}), for the same models of Fig.~\ref{fig:tau_spec} (surface temperature indicated in the legend).}
\label{fig:tau}
\end{figure}

Since in the analytical models of the OG the energies of $\gamma$-ray photons are fixed to a single value, it is useful to look at the energy for which the specific optical depth is the largest. For $E_\gamma=1$ GeV (blue lines), the specific $\tau$ rises sharply and the maximum lies at $\lesssim 2 E_{\rm min}$ (the latter being the energy for which the sharp drops begins). For larger $\gamma$-ray energies, the energy-dependence of the specific optical depth is less pronounced.

The integrated optical depth, Eq.~(\ref{eq:tau}), depends on the spectral distribution of the thermal photons. In Fig.~\ref{fig:tau} we compare $\tau$ for the models shown in Fig.~\ref{fig:tau_spec}. The larger the temperature, i.e., the X-ray flux, the larger the maximum $\tau_{\gamma\gamma}$. Note that the optical depth scales with the position of the gap, and with the size of the emitting region, $r_{bb}$. As a consequence, for a fixed temperature, the optical depth can be several orders of magnitude smaller or larger. Regardless of that, however, we can generally identify the energy where $\tau_{\gamma\gamma}$ peaks (Fig.~\ref{fig:tau}):

\begin{equation}\label{eq:egamma_taumax}
  E_\gamma(\tau_{\rm max}) \sim \frac{0.25}{kT {\rm [keV]}} {\rm GeV}~.
\end{equation}
As discussed in Paper I, the outer gap models rely on the estimate of the typical energy of the interacting photons. Since the cross-section depends on the product $E_XE_\gamma$, one can introduce a parameter $k_\gamma$ to estimate the most likely interacting $\gamma$-ray photon, for a given X-ray flux given by a blackbody distribution:
\begin{equation} \label{eq:egamma}
E_\gamma = \frac{k_\gamma}{kT [{\rm keV}]} {\rm ~GeV} ~.
\end{equation}
The value of $k_\gamma$ implicitly includes the choices of the typical energies of both photons (see \S~2.5 of Paper I). Given the optical depth shown in Fig.~\ref{fig:tau}, and Eq.~(\ref{eq:egamma_taumax}), we consider $k_\gamma \sim 0.1-0.3$, corresponding to the energies of $\gamma$-rays where the optical depth is larger. Note that, instead, \cite{zhang97} consider an X-ray photon with energy $E_X=3~kT$, and estimate the minimum required $\gamma$-ray energy to interact with it, i.e., for which the cross section vanishes, Eq.~(\ref{eq:pp_cross_section}). In other words, they consider $E_\gamma = (m_ec^2)^2/3kT$, i.e., $k_\gamma=0.087$, a value for which the optical depth is relatively small (see Fig.~\ref{fig:tau}).

\section{The $\gamma$-ray Spectrum}\label{sec:spectrum}

\subsection{Losses per particle}

In the OG model, the $\gamma$-ray radiation is thought to be produced by a population of relativistic pairs spiraling around, while sliding, curved magnetic field lines.
These particles produce synchro-curvature radiation. \cite{cheng96} derive and discuss the formulae describing the energy loss for one particle, and \cite{paper0} have recently simplified these to the following:\footnote{The expression for the $\gamma$-ray spectrum given by equation (57) in \citet{zhang97}, derived from the original formula of \cite{cheng96}, has a few typos, which have been taken over to several other works quoting it. In that equation $r_{\rm eff}$ should be $r_{\rm eff}^2$ when it is multiplying $Q_2^2$ and the factor $y$ multiplying only $K_{2/3}(y)$ should multiply the whole square bracket of Eq.~(\ref{eq:sed_x}).}

\begin{equation}\label{eq:sed_synchrocurv}
 \frac{dP_{sc}}{dE_\gamma} = \frac{\sqrt{3} e^2 \Gamma y}{4\pi \hbar r_{\rm eff} } [ (1 + z) F(y) - (1 - z) K_{2/3}(y)]~,
\end{equation}
where
\begin{eqnarray}
 && F(y) = \int_y^\infty K_{5/3}(y') dy'~,\label{eq:f_y}\\
 && y=\frac{E}{E_c} ~,\\
 && E_c = \frac{3}{2}\hbar cQ_2\Gamma^3~,\label{eq:echar}
\end{eqnarray}
where $E_c$ is the characteristic energy, $K_n$ are the modified Bessel functions of the second kind of index $n$, and
\begin{eqnarray}
 && Q_2 = \frac{\cos^2\alpha}{r_c}\sqrt{1 + 3\xi  + \xi^2 + \frac{r_{\rm gyr}}{r_c}} \label{eq:q2}~, \\
 && r_{\rm gyr} \equiv \frac{mc^2\Gamma\sin\alpha}{eB}~,\\
 && \xi \equiv \frac{r_c}{r_{\rm gyr}}\frac{\sin^2\alpha}{\cos^2\alpha}~,\\
 && z= (Q_2 r_{\rm eff})^{-2} ~,\\
 && r_{\rm eff} = \frac{r_c}{\cos^2\alpha}\left(1 + \xi+ \frac{r_{\rm gyr}}{r_c}  \right)^{-1}~,\\
 && g_r =  \frac{r_c^2}{r_{\rm eff}^2}\frac{[1 + 7(r_{\rm eff}Q_2)^{-2}]}{8 (Q_2r_{\rm eff})^{-1}}~.\label{eq:gr}
\end{eqnarray}
Here, $r_c$ is the radius of curvature of their trajectory, $r_{\rm gyr}$ is the Larmor radius, and $r_{\rm eff}$, $Q_2$, $\xi$, and $g_r$ are factors introduced to provide a compact formula and explicitly distinguish the limiting cases. The peak of the spectrum is located close to $E_c$ and, for energies $E\ll E_c$, the dominant term in Eq.~(\ref{eq:sed_synchrocurv}), $F(y)$, provides a precise value for the low energy (low-energy) slope spectrum behavior, $dP_{sc}/dE \sim E^{0.25}$. In \cite{paper0} we describe in which case Eq.~(\ref{eq:sed_synchrocurv}) reduces to synchrotron ($\xi\gg1$) or curvature radiation ($\xi\ll 1$) formulae. 

In order to calculate the observed spectrum, one has to consider the radiation emitted towards the Earth by the particles which move along the gap. Let us indicate with $dN/dx$ the number of such particles per unit of distance $x$ along the line. This {\em effective particle distribution} implicitly takes into account the geometry and beaming effects (i.e., it only considers  particles which emit radiation pointing toward the Earth) and can, in principle, differ from the total particle distribution which may never be inferred.

The total spectrum is obtained by integrating the single-particle spectrum, Eq.~(\ref{eq:sed_synchrocurv}), along the traveled distance between the assumed boundaries $x_{\rm in}$ and $x_{\rm out}$, convoluted with the effective particle distribution:

\begin{eqnarray}\label{eq:sed_x}
  \frac{dP_{\rm gap}}{dE_\gamma} & = &  \int_{x_{\rm in}}^{x_{\rm out}} \frac{dP_{sc}}{dE_\gamma}\frac{dN}{d x} {\rm d}x~\nonumber \\
 & = & \frac{\sqrt{3} e^2}{2 h}  y \int_{x_{\rm in}}^{x_{\rm out}} \frac{dN}{dx} \frac{\Gamma}{r_{\rm eff}}\times\\ 
 && [ (1 + z) F\left(y\right) - (1 - z) K_{2/3}(y)] dx~. \nonumber
\end{eqnarray}
In Eq.~(\ref{eq:sed_x}), the values of the quantities $Q_2$, $r_{\rm eff}$, $E_c$, $dN/dx$ depend, in general, on the local values of $\Gamma$, $\sin\alpha$, $r_c$, $B$, and $E_\parallel$. Below we discuss the dependence of $r_c$, $B$, $E_\parallel$, $dN/dx$ on $x$, and the particle dynamics along the gap (evolution of $\Gamma$ and $\sin\alpha$).

It is also interesting to note the difference of our approach to that used for instance by \cite{zhang07}. First, we have already commented in Paper I about the changes introduced by the proper consideration of uncertainties in many of the model parameters (which were before assumed to adopt a single, fixed value). But in addition, we are here (see below) considering the solution for the particle dynamics along the gap, including the full synchro-curvature effects (whereas only curvature was studied earlier), and in particular, the initial synchrotron-dominated losses.

This allows us to consistently compute the evolution of the main quantities of the model (radius of curvature, magnetic field) along the gap, and then to compute the spectra using these values. Finally, we note that in \cite{zhang07} the effective particle distribution is assumed as $dN/dx \sim r_c^{9/4}$ whose normalization and shape is considered to come from the counting of the pairs produced through the different particles. In Paper I, we have demonstrated that these precise quantitative predictions rely on some underlying rough approximations and assumptions, like the dependences on $B$ and $P$. In particular, \cite{zhang97} fix the value of the pitch angle (their Eq.~51), which is not physically justified (see \S~\ref{sec:dynamics} and \citealt{paper0}), which lead to a particular dependence with $x$. However, we do not expect $dN/dx$ to have such fixed dependence with $x$, thus we parametrize it and study different alternatives.

Compared with the synchro-curvature spectra calculated in \cite{paper0}, we note two main differences. First, in that work we calculate the average of the radiation emitted by a single particle along its trajectory, considering different relative weights assigned to different positions. Such weights mimic the geometry and viewing angle effects, and are implicitly included in the parametrization of $dN/dx$ of this work. Second, in \cite{paper0} we considered constant values of $B$, and $r_c$, contrary to what is done here, where such quantities vary along the gap.

\subsection{Geometry and electromagnetic fields}

The magnetospheric geometry and the position of the gap are not well-known (a dipole can only be a rough approximation in the outer gap, see Paper I), but the basic length-scale of the problem is the light cylinder, which depends on the spin period:
\begin{equation}
R_{lc} = \frac{Pc}{2\pi} \simeq 4.77\times 10^9 P{\rm[s] ~ cm} ~.\label{eq:rlc}
\end{equation}
The OG is thought to extend between the null surface and an outer boundary close to the light cylinder. As a consequence, and as we have seen in Paper I, realistic values of $x/R_{lc}$ are between 0.3 and a few. Within this range, we can consider it as a free parameter, with the interval boundaries $x_{\rm in},x_{\rm out}$.

Since the integration variable of Eq.~(\ref{eq:sed_x}) is the distance along the line, $x$, we have to express $r_c(x)$, $B(x)$ and $E_\parallel(x)$. Given the uncertainties (see the discussion in Paper I), we parametrize the radius of curvature as 
\begin{equation}
 r_c = R_{lc}\left(\frac{x}{R_{lc}}\right)^\eta~.
\end{equation}
Close to the surface, $\eta \sim 0.5$, then the relation is less trivial, so we will take such power law as an effective way to model our ignorance, with $\eta \sim 0.2-1$.

The functional form $B(x)$ is discussed in detail in the appendix of Paper I. Here we briefly summarize the main considerations. The surface dipolar field $B_\star$ (at the pole) is observationally estimated by assuming that the electromagnetic torque regulates the period evolution:
\begin{equation}\label{eq:inferred_bpole}
 B_\star = 6.4 \times 10^{19} ~ \sqrt{P{\rm [s]}\dot{P}}~\mbox{G}~,
\end{equation}
where the uncertainties on the moment of inertia, inclination angle and radius of the NS make such estimate accurate within a factor of $\sim 2$. The distance along the line can be roughly taken to be of the order of the distance from the surface, $x\sim r$, and, assuming a simple power-law for the radial dependence $B(r) \sim B_\star (R_\star/r)^b$, we have that
\begin{equation}\label{eq:b_xc}
  B(x) = B_\star \left(\frac{R_\star}{x}\right)^b~.
\end{equation}
The numerical solutions for the most simple magnetospheric configuration of a pulsar \citep{contopoulos99} can be approximated by a vacuum dipole close to the surface, $B\propto r^{-3}$, and by a split monopole $B\propto r^{-2}$, close to and beyond the light cylinder. In the region of interest, the geometry of a realistic magnetosphere presents more stretched lines and softer radial decay, compared to a dipolar field, thus we explore $b$ in the range 2-3 (see Paper I for further details).

Regarding the local value of $E_\parallel$, the Poisson equation, solved in the vacuum thin gap limit under several approximations \citep{cheng86a,paper1}, predicts a proportionality with the magnetic flux across the gap. Since this is constant by construction (the gap is supposed to be confined by two magnetic field lines), $E_\parallel$ is supposed to be constant along the magnetic field line. This has been roughly confirmed by numerical calculations for thin gaps \citep{hirotani06}, but it may not apply when screening effects (given by spatial-dependent pair production) and/or a non-negligible thickness are considered. However, due to the lack of adequate formalism able to find more general solutions for $E_\parallel$, here we consider a constant value of $E_\parallel$, treating such value as a model parameter. Below, we explore different values of $E_\parallel \sim 10^6-10^8$ V/m.

A way to constrain this value is, in principle, to impose the gap closure, as done in the thick OG. \cite{zhang97} fix the characteristic energy, Eq.~(\ref{eq:echar}), by requiring that the synchro-curvature characteristic energy, $E_c$, coincides with the most likely interacting $\gamma$-rays, as discussed in Paper I and in \S\ref{sec:pair_production} above. Thus, once $E_c$ has been fixed, the gap is closed and $E_\parallel$ is estimated as (Paper I)

\begin{eqnarray}
  E_\parallel & =& \left(\frac{2}{3}\right)^{7/3}\frac{e}{c^2}\left(\frac{E_\gamma}{\hbar}\right)^{4/3}\frac{g_r}{(Q_2r_c)^{4/3}}\left(\frac{2\pi R_{lc}}{P r_c}\right)^{2/3} \nonumber\\
  &\simeq & 3.6\times 10^6  \frac{g_r R_{lc}^{2/3}}{Q_2^{4/3}r_c^2 P{\rm [s]}^{2/3}} \left(\frac{k_\gamma}{kT [{\rm keV}]}\right)^{4/3} \frac{{\rm V}}{{\rm m}}~. \label{eq:epar_t}
\end{eqnarray}
Relying on this formula, one can establish a possible relation to the surface temperature, keeping in mind the underlying uncertainties and spatial-dependent range of values of the parameters appearing in Eq.~(\ref{eq:epar_t}), as discussed in Paper I. Note, however, that this method, consisting in fixing a single value $E_c$, is equivalent to fix the shape of the synchro-curvature spectrum, given by the single-particle spectrum. This limitation does not allow us to reproduce the observed variety of shapes of $\gamma$-ray spectra (see \S\ref{sec:results}).

\subsection{Particle dynamics}\label{sec:dynamics}

\begin{figure}
\centering
\includegraphics[width=.5\textwidth]{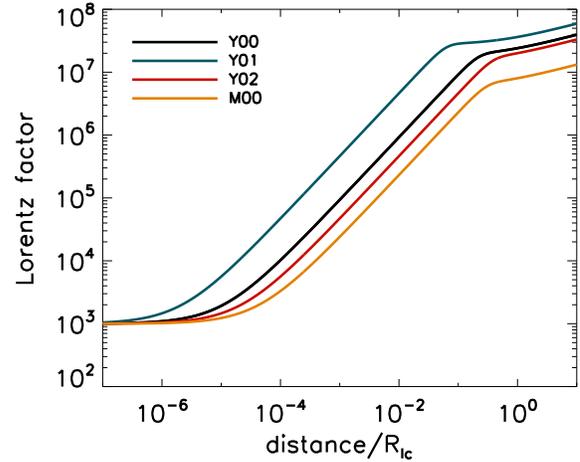}
\caption{Evolution of $\Gamma$ for a charged electron/positron, as a function of the distance travelled along a magnetic field line (normalized by $R_{lc}$, and where the origin is the place of particle creation), for the models Y00, Y01, Y02 and M00 of Table~\ref{tab:spe_models}.}
 \label{fig:tra2d_gap}
\end{figure}

In order to obtain the values of $\Gamma$ and $\sin\alpha$ along the trajectory, we simulate the motion of charged particles along the line. In \cite{paper0} we already did it for a fixed set of values for $r_c,E_\parallel,B$. We set the initial values for the Lorentz factor $\Gamma_{\rm in}=10^3$ (a typical value for a pair created by a GeV photon) and the pitch angle, $\sin\alpha_{\rm in}=0.5$ (representing a uniformly chosen random value, which makes sense if the mean free path of $\gamma$-ray photons is large enough, otherwise $\sin\alpha_{\rm in}\ll 1$). Then we evolve the parallel and perpendicular momenta of particles according to the equations of motion \citep{paper0}:

\begin{eqnarray}
 && \frac{\de(p\sin\alpha)}{\de t} = - \frac{P_{sc}\sin\alpha}{v}~, \label{eq:motion_perp} \\
 && \frac{\de(p\cos\alpha)}{\de t} = eE_\parallel - \frac{P_{sc}\cos\alpha}{v}~. \label{eq:motion_par}
\end{eqnarray}
where $p=\Gamma m v$ is the momentum of the particle, $v$ its spatial velocity, and $P_{sc}$ is the synchrotron power radiated by one particle, obtained by integrating in energy Eq.~(\ref{eq:sed_synchrocurv}):

\begin{equation}\label{eq:power_synchrocurv}
 P_{sc} = \frac{2e^2 \Gamma^4 c}{3 r_c^2} g_r~.
\end{equation}
We show the evolution of $\Gamma$ in Fig.~\ref{fig:tra2d_gap}, for four of the models of Table~\ref{tab:spe_models}. Soon after pair creation, $\xi \gg 1$, i.e. the losses mostly regard the perpendicular momentum, being approximated by synchrotron emission. Then, $\Gamma$ increases and $\sin\alpha$ becomes very small. After having traveled a distance $(x-x_{\rm in}) \sim 0.01-1 R_{lc}$, the radiative losses can be approximated by purely curvature radiation. In this regime, electrons are mono-energetic at a given point in the gap, because their Lorentz factor $\Gamma$ is determined according to the radiation-reaction steady state (i.e., electric force $eE_\parallel$ balanced by purely curvature losses):
\begin{equation}
 \Gamma_{\rm eq} = \left( \frac{3}{2}\frac{E_{\parallel} r_c^2}{e} \right)^{1/4}~. \label{eq:gamma}
\end{equation}
Results are very similar to \cite{paper0} and \cite{hirotani99a}, with the difference being that the equilibrium value of  $\Gamma$ varies with the position, due to the change of the local values of $r_c$ and $B$. Since we assume $E_\parallel$ to be constant, $r_c$ increases and $B$ decreases, then the value of $\Gamma_{\rm eq}$ slightly increases. On the other hand, we find the same trends: the larger the values of $E_\parallel$ and/or $r_c$, the larger is $\Gamma$, because of relatively inefficient radiative losses, as seen in Eq.~(\ref{eq:gamma}).

In general, particles are able to quickly reach such equilibrium state only if $E_\parallel$ and $r_c$ are large enough. However, in  general, in the initial part of the trajectory the pitch angle is not negligible, the Lorentz factor is smaller, and the emitted radiation can be important. This is why we need to consider the full expression of synchro-curvature radiation and evaluate it along the whole trajectory of the particles moving through the gap.

\begin{figure}
\centering
\includegraphics[width=.5\textwidth]{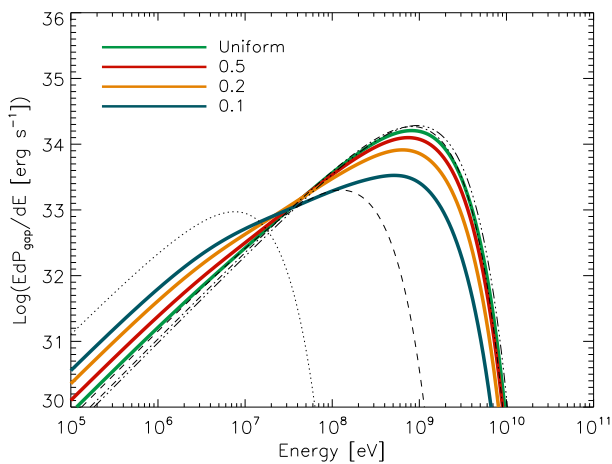}
\caption{Gap-integrated spectra for the model Y00 of Table~\ref{tab:spe_models}. In solid colored lines we show the trajectory-integrated spectra for different effective particle distribution (uniform in green or exponential, Eq.~\ref{eq:n_xc}, with $x_0/R_{lc}=0.5,0.2$ or 0.1 in red, orange and blue, respectively). Dotted, dashes, dot-dashed, triple dot-dashed lines indicate the single-particle spectra emitted at the specific positions $(x-x_{\rm in})/R_{lc}=0.01,0.1,0.5,1.0$, respectively, multiplied by $N_0B_{12}/P$, in order to be visually comparable with the gap-integrated values.}
 \label{fig:spe_gap_comp}
\end{figure}

\begin{table*}
\begin{center}
\caption{Models used for the spectrum calculation: input parameters, peak of the $E^2dN/dE$ distribution, $\gamma$-ray efficiency (luminosity in 0.1-100 GeV over rotational energy). We distinguish between young pulsar (Y) and millisecond pulsar (M) models. We consider two baseline models, M00 and Y00, and the other models consider the variation of a single parameter. See Fig.~\ref{fig:spectra} for the expected spectra.}
\label{tab:spe_models}
\begin{tabular}[ht!]{c c c c c c c c c c c c c c c c c}
\hline
\hline
Model & $P$[s]  & $x_{\rm in}/R_{lc}$ & $x_{\rm out}/R_{lc}$ & $\eta$ & $\log(B_\star {\rm [G]})$ & $b$ & $E_\parallel$[$10^7$ V/m] & $E_{\rm peak}$[GeV]  & & $L_\gamma/\dot{E}_{\rm rot}$ & \\
& & & & & & & $w=1$ & $\frac{x_0}{R_{lc}}=0.1$ & $w=1$ & $\frac{x_0}{R_{lc}}=0.1$ & $w=1$ \\
\hline
Y00 & 0.1	 & 0.5 & 1.5 &	0.5	& 12	& 2.5	  & 1  & 0.83 & 0.51 & 0.38 & 0.082 \\
YE1 & 0.1	 & 0.5 & 1.5 &	0.5	& 12	& 2.5	  &0.5& 0.45 & 0.16 & 0.13 & 0.096 \\
YE2	& 0.1	 & 0.5 & 1.5 &	0.5	& 12	& 2.5	  & 2  & 1.45 & 0.89 & 0.89 & 0.37 \\
YE3	& 0.1	 & 0.5 & 1.5 &	0.5	& 12	& 2.5	  & 3  & 1.91 & 1.35 & (1.41) & 0.78 \\
YB1	& 0.1	 & 0.5 & 1.5 &	0.5	&12.3& 2.5 & 1  & 0.83 & 0.51 & 0.075 & 0.016 \\
YB2	& 0.1	 & 0.5 & 1.5 &	0.5	& 13	& 2.5	  & 1  & 0.83 & 0.51 & 0.038 & 0.008 \\
YB3	& 0.1	 & 0.5 & 1.5 &	0.5	& 12	& 3	  & 1  & 0.83 & 0.55 & 0.38 & 0.078 \\
YB4	& 0.1	 & 0.5 & 1.5 &	0.5	& 12	& 2	  & 1  & 0.83 & 0.51 & 0.38 & 0.083 \\
YN1 & 0.1	 & 0.5 & 1.5 &	0.2	& 12	& 2.5	  & 1  & 0.83 & 0.55 & 0.38 & 0.077 \\
YN2 & 0.1	 & 0.5 & 1.5 &	1.0	& 12	& 2.5	  & 1  & 0.83 & 0.44 & 0.36 & 0.091 \\
YX1 & 0.1	 & 0.3 & 1.5 &	0.5	& 12	& 2.5	  & 1  & 0.78 & 0.48 & 0.38 & 0.090 \\
YX2 & 0.1	 & 1.0 & 1.5 &	0.5	& 12	& 2.5	  & 1  & 0.67 & 0.63 & 0.20 & 0.15 \\
YX3 & 0.1	 & 0.5 & 1.0 &	0.5	& 12	& 2.5	  & 1  & 0.83 & 0.51 & 0.38 & 0.082 \\
YX4 & 0.1	 & 0.5 & 2.0 &	0.5	& 12	& 2.5	  & 1  & 0.83 & 0.51 & 0.38 & 0.082 \\
YP1 & 0.05& 0.5 & 1.5 &	0.5	& 12	& 2.5	  & 1  & 0.55 & 0.26 & 0.033 & 0.004 \\
YP2 & 0.2	 & 0.5 & 1.5 &	0.5	& 12	& 2.5	  & 1  & 0.78 & 0.51 & 0.26 & 0.10 \\
\hline
M00 & 0.005 & 0.5 & 1.5 & 0.5	& 8	& 2.5	 & 5  & 0.59 & 0.29 & (1.73) & 0.24 \\
ME1 & 0.005 & 0.5 & 1.5 & 0.5	& 8	& 2.5	 & 1   & 0.042 & 0.007 & 0.0034 & 0.00004 \\
ME2 & 0.005 & 0.5 &	 1.5 & 0.5	& 8	& 2.5	 & 10 & 1.02 & 0.78 & (4.57) & (1.47) \\
\hline
\hline
\end{tabular}
\end{center}
\end{table*}

\begin{figure*}
\centering
\includegraphics[width=.5\textwidth]{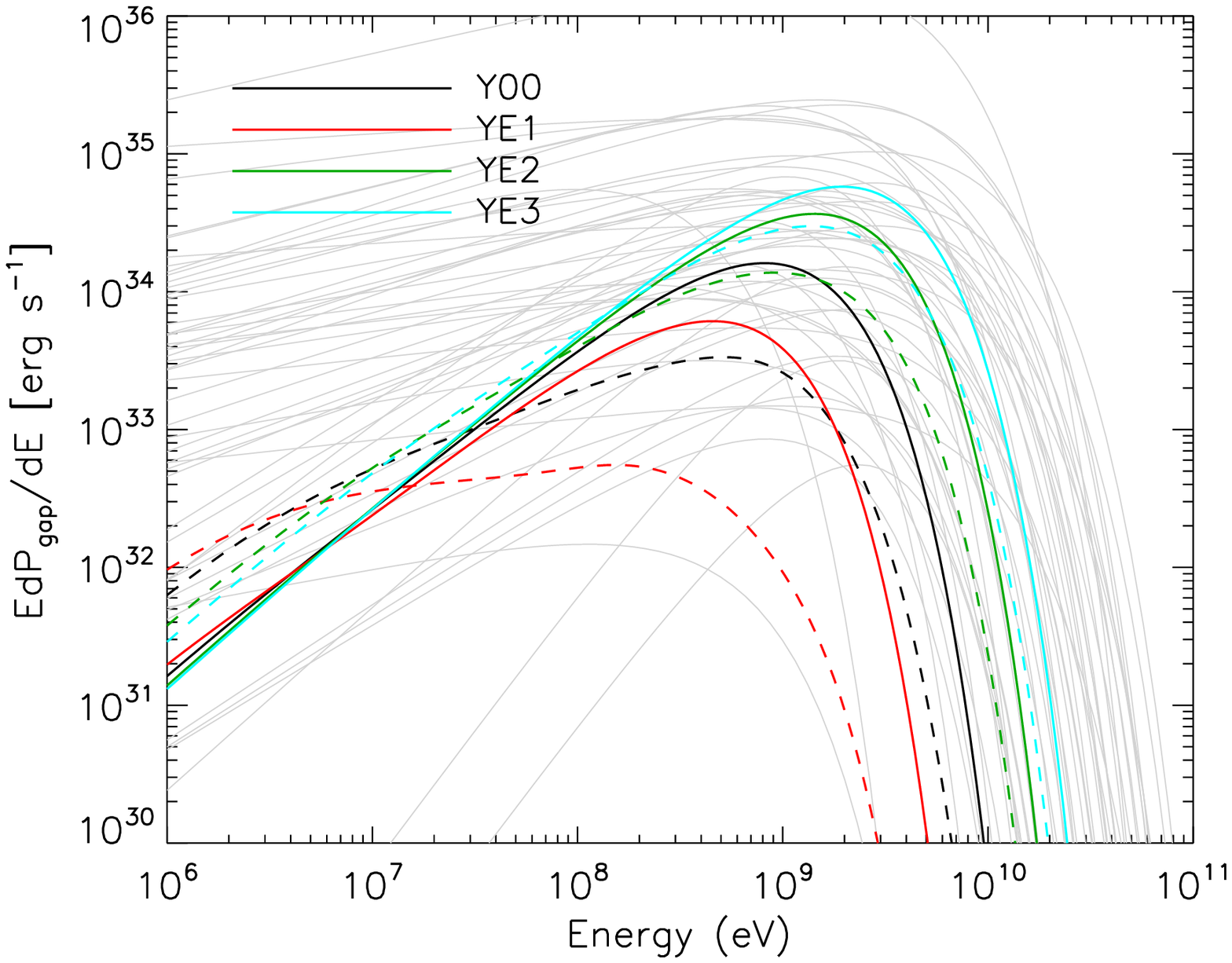} \hspace{-1cm}
\includegraphics[width=.5\textwidth]{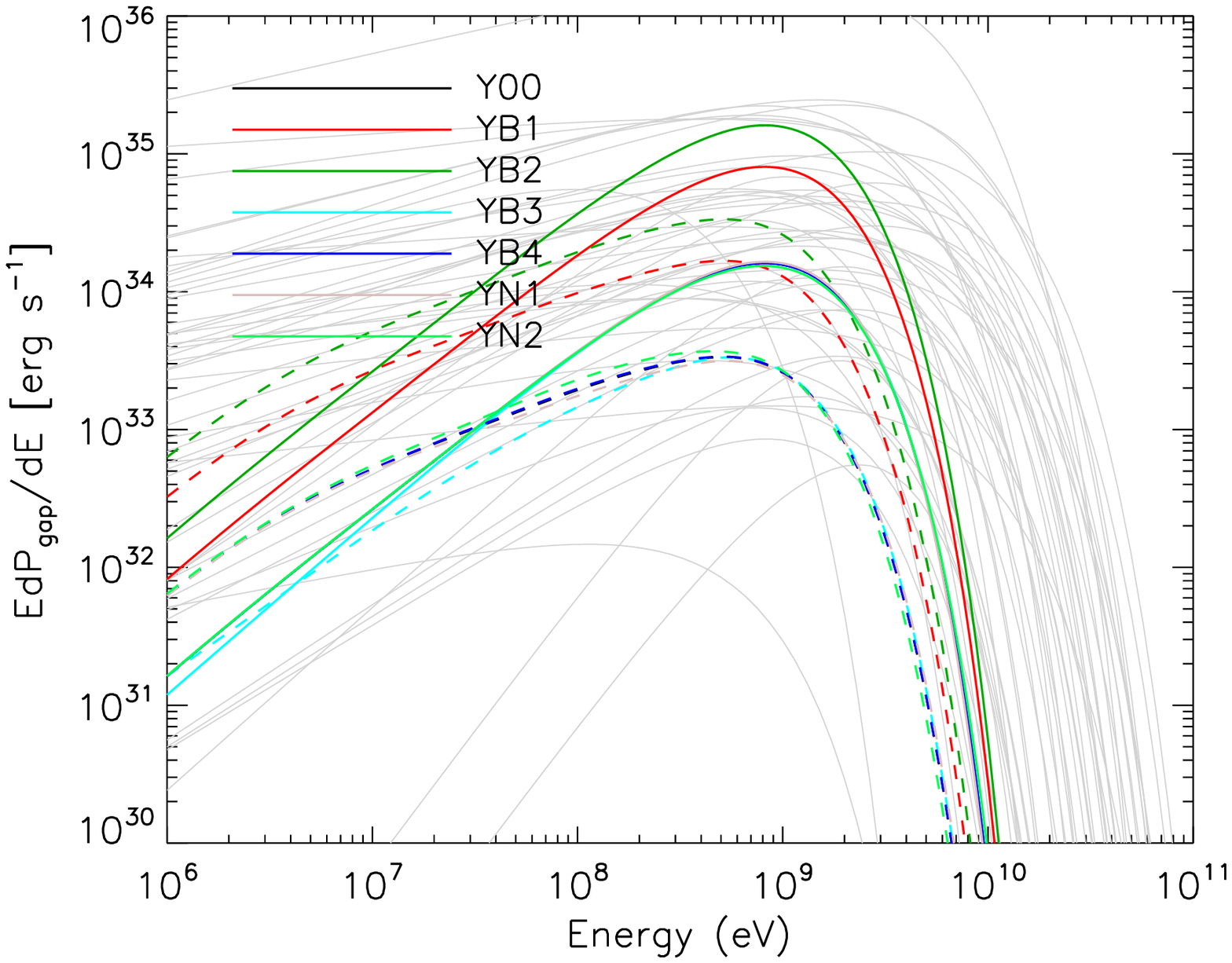}\\
\includegraphics[width=.5\textwidth]{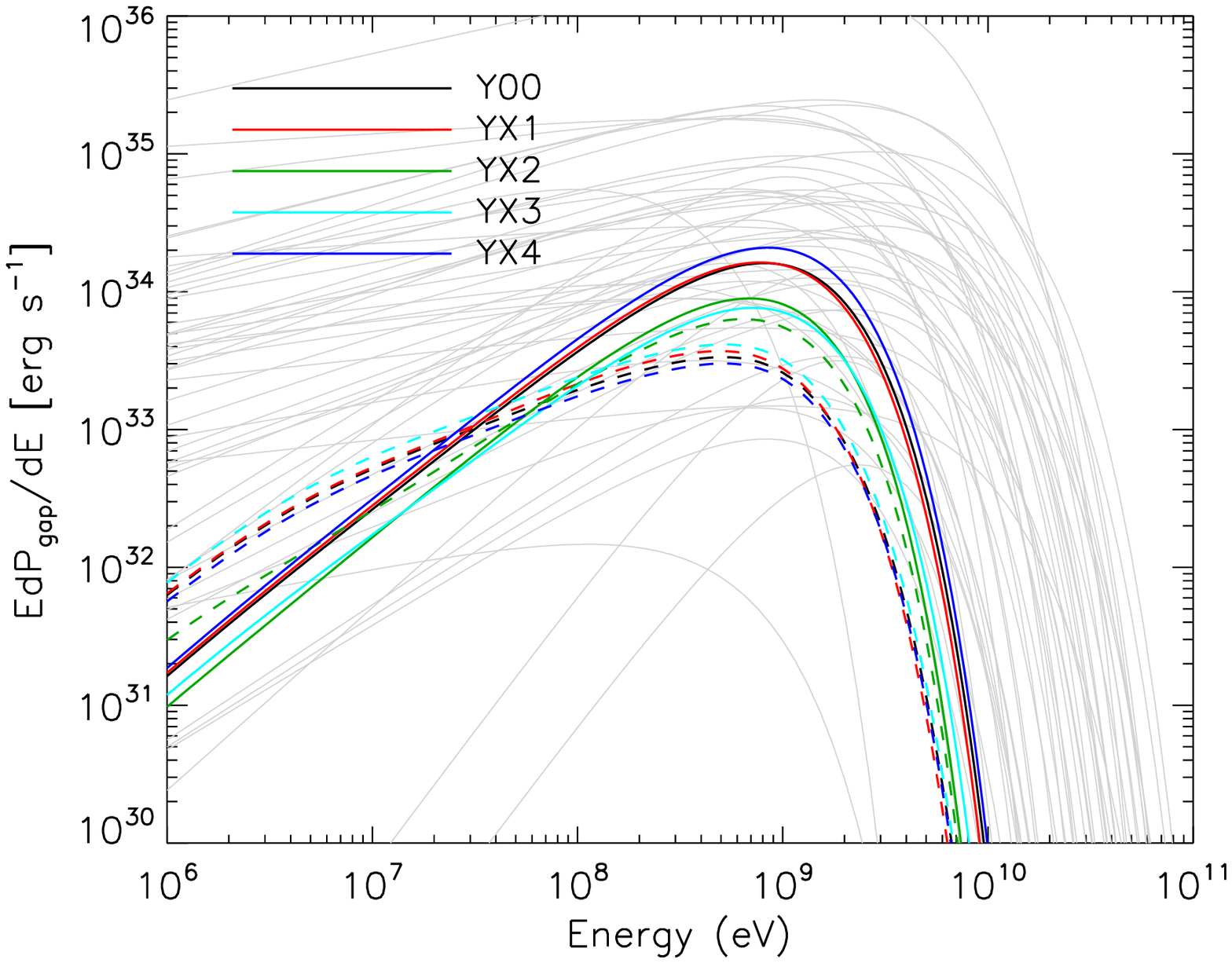}\hspace{-1cm}
\includegraphics[width=.5\textwidth]{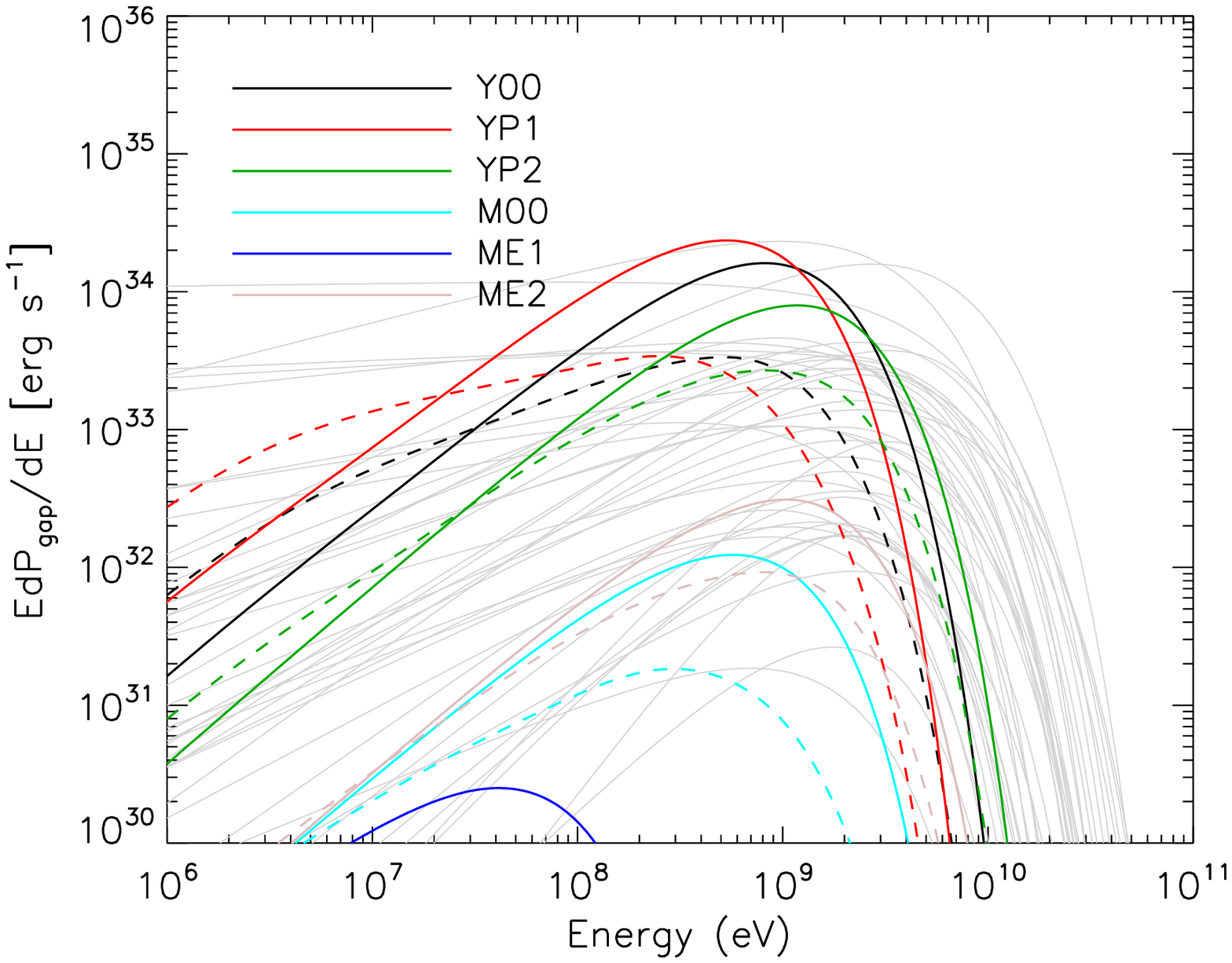}
\caption{$\gamma$-ray spectra obtained for different gap models of Table~\ref{tab:spe_models}. Solid curves are for uniform $dN/dx$, while dashed lines are for the exponential distribution with $x_0=0.1 R_{lc}$. Left top panel: changes in $E_\parallel$; top right: changes in $B_\star$, $b$ or $\eta$; bottom left: changes in $x_{\rm in}$ or $x_{\rm out}$; bottom right: changes in spin period. The gray lines are the set of cut-off power law that best fit the observed spectra in the range 0.1-100 GeV, as taken from the {\em Fermi}-LAT catalog, and extrapolated here down to 1 MeV. They represent all the young pulsars ($P>10$ ms), except the bottom right panel, where we show the millisecond sample only.}
 \label{fig:spectra}
\end{figure*}

\subsection{Effective particle distribution}

In order to predict the observed spectra, we need to convolve the single-particle spectrum along the trajectory with the effective particle distribution, i.e., particles which emit high-energy radiation towards us. \cite{goldreich69} and all the force-free magnetosphere models estimate the particle density with the Goldreich-Julian value, given by
\begin{equation}
  n_{gj} = -\frac{\vec{\Omega}\cdot\vec{B}}{2\pi ec[1-(r/R_{lc})^2]}~,\label{eq:n_gj}
\end{equation}
However, in a gap, for definition, there is a lack of particles. The $e^--e^+$ pairs are continuously created and accelerated away, thus partially screening the parallel electric field. A precise value of the particle density requires self-consistent numerical simulations. Moreover, not all the particles will emit radiation towards us. As a consequence, depending on the magnetospheric geometry and viewing angles, the effective particle distribution differs from the total particle distribution in the gap. Since neither the total nor the effective number of particles are a-priori easy to estimate, we assume that the gap-integrated effective number of particles is $N_0B_{12}/P{\rm [s]}$, where $N_0$ is a free parameter and the scaling $B/P$ is motivated by Eq.~(\ref{eq:n_gj}). In conclusion, we parametrize the effective particle distribution as follows:
\begin{equation}\label{eq:n_xc}
 \frac{dN}{dx}=N_0\frac{B_{12}}{P{\rm [s]}}w(x)=N_0\frac{B_{12}}{P{\rm [s]}}\frac{e^{-(x-x_{\rm in})/x_0}}{x_0(1 - e^{-x_{\rm out}/x_0}) }~,
\end{equation}
where we have introduced a relative weight, $w(x)$, which depends on the length-scale $x_0$. The arbitrarily chosen functional form (an exponential distribution) is one of the infinite possible, effective ways to mimic possible effects of geometry (position of the gap, viewing angle and beaming), and, if $x_0/R_{lc}\lesssim 1$, to assign more importance to the radiation emitted in the initial part of the trajectory, rather than the final. Other functional forms with similar relative weights between the inner and outer parts of the trajectory would produce the same results.

A large weight to the initial part of the trajectories can be physically, qualitatively justified by a number of effects, as discussed also in \cite{paper0}, to which we refer for further details. A trivial one is a favorable viewing angle. Second, the cone of the emitted radiation at low $\Gamma$ and low photon energies is larger, thus we expect the beaming angle to vary along the trajectories. Third, here we are considering only the primary particles, but a cascade of secondary and tertiary particles are expected to flow with lower Lorentz factor \citep{cheng86b}. Such cascade would lose mostly perpendicular momentum, providing synchrotron-like radiation. Of course, the particle distribution above is an effective way to account for these effects; a quantitative characterization of its functional form could only be indicated by numerical simulations that consider the complex production, interaction, and propagation of photons and particles.

Finally, note that the same exponential distribution was applied in our previous study of the radiation emitted by a single particle along its trajectory, with constant $B$ and $r_c$ \citep{paper0}. In that case, we studied different weights, with normalization equal to one, to get the average of emitted spectra along the trajectory. The variety of expected spectra is similar to the one we expose below.

\section{Results}\label{sec:results}

According to the setup of the model described above, for a pulsar of given period $P$ (which defines $R_{lc}$, Eq.~\ref{eq:rlc}), and surface magnetic field $B_\star$, the free parameters are: $x_{\rm in}$, $x_{\rm out}$, $\eta$, $b$, $E_\parallel$, and the particle distribution (uniform or, e.g. in our setting, an exponential distribution with length-scale $x_0$). We fix the remaining parameters: $N_0=10^{30}$ (typical value for the Goldreich-Julian density integrated through the volume of the gap, e.g., \citealt{zhang97}), $x_{\rm out} = x_{\rm in} + R_{lc}$ (i.e., we fix the total length of the gap to be $R_{lc}$), $\Gamma_{\rm in}=10^3$, and the initial pitch angle $\sin\alpha_{\rm in}=0.5$.

In Table~\ref{tab:spe_models}, we consider models with different sets of parameters. We divide them in two families: young pulsars (Y), and millisecond pulsars (M), according to the period, which sets the light cylinder distance, the typical length-scale of the problem. We take as baseline models Y00 and M00, and we consider the other models by changing one of the parameters: $E_\parallel$ (models for which labels have a letter E), $B_\star$ or $b$ (B), $\eta$ (N), gap extremes (X), period (P). This representative, incomplete sample of possible models allows to see general trends in the large space of parameters.

In Fig.~\ref{fig:spe_gap_comp} we show the expected spectrum for the base model for young pulsars, Y00 (see Table for the values). With black dashes, we show the single-particle spectra (multiplied by $N_0$) at different positions along the trajectory (from left to right: $x/R_{lc}=0.01,0.1,0.5,1.0$, respectively). With colors, we plot the spectra for different effective particle distributions (uniform or exponential with $x_0/R_{lc}$ indicated in the legend). The higher the number of photons coming from the initial part of the trajectory (i.e., $x_0\ll R_{lc}$), the lower is the peak of the spectral energy distribution. As a consequence, the low-energy slope can be much softer than that resulting from the synchro-curvature spectrum calculated for the external, steady-state regime of the trajectory, as usually done in literature. On the other hand, distributions with larger $x_0$ do not show any important difference with the synchro-curvature spectra computed for values at steady-state (e.g., approximated by curvature radiation).

The same trend is seen in all models, which are plotted in Fig.~\ref{fig:spectra}. We see that some parameters produce important changes: in particular, the values of $E_\parallel$, $R_{lc}$ and $B_\star$, besides the already discussed importance of the particle distribution. On the other hand, the values of $\eta$ and $b$ play a lesser role, at least within the explored range. For millisecond pulsar models, we need a larger $E_\parallel$ to explain their luminosity. This is consistent with the fact the classical OG model, when applied to such sources, needs the presence of strong multipolar components in the magnetic field, which eventually can support a larger electric field \citep{zhang03}.

We note that previous works, see \cite{zhang97} and thereafter, do not find any low-energy slope change despite they have also considered relative weights of the spectrum (although see caveats above) because they only considered radiation emitted in the reaction limit saturation; neglecting the initial synchrotron dominated regime in the particle dynamics.

\begin{figure*}
\centering
\includegraphics[width=.5\textwidth]{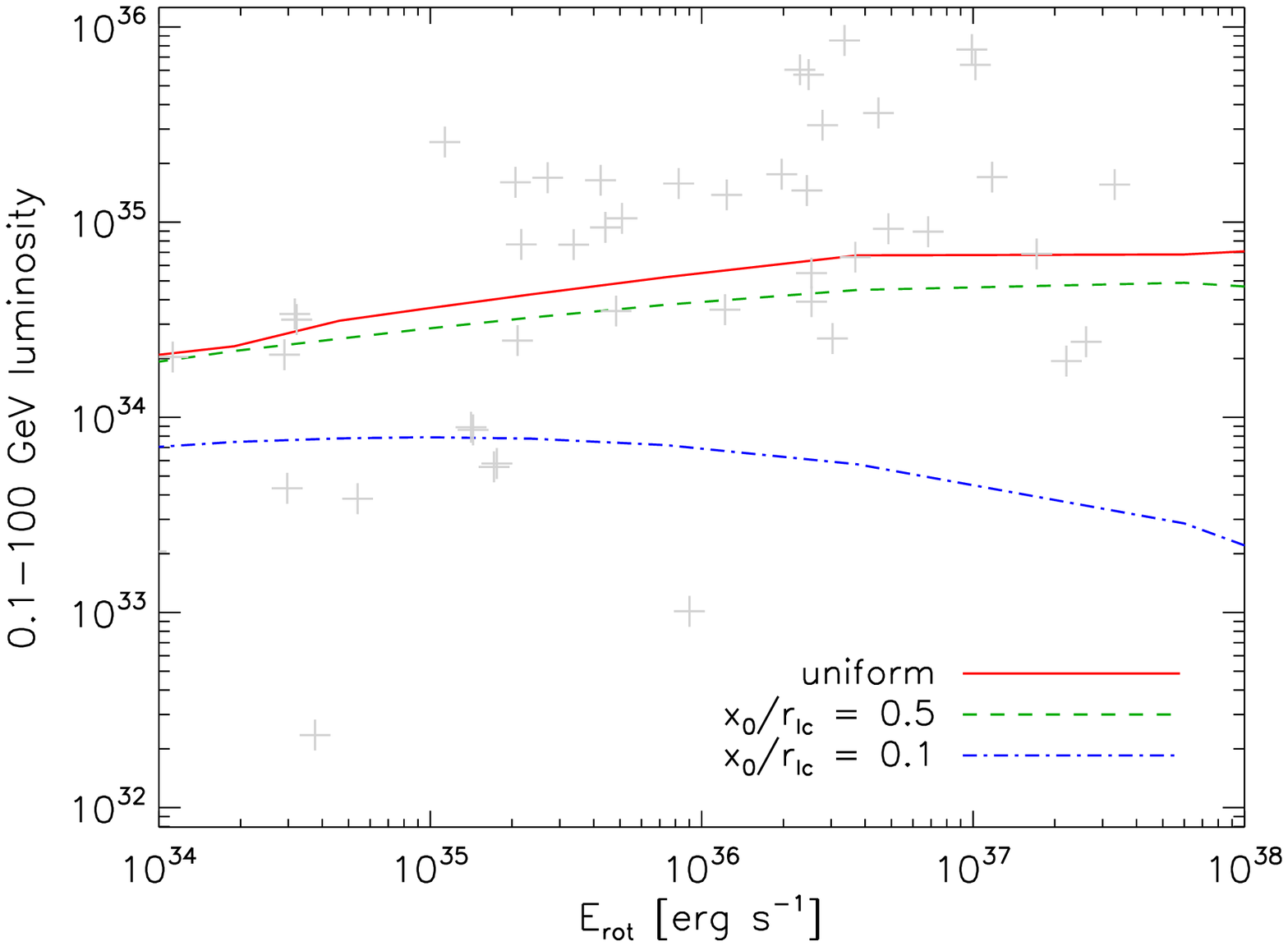} \hspace{-.7cm}
\includegraphics[width=.5\textwidth]{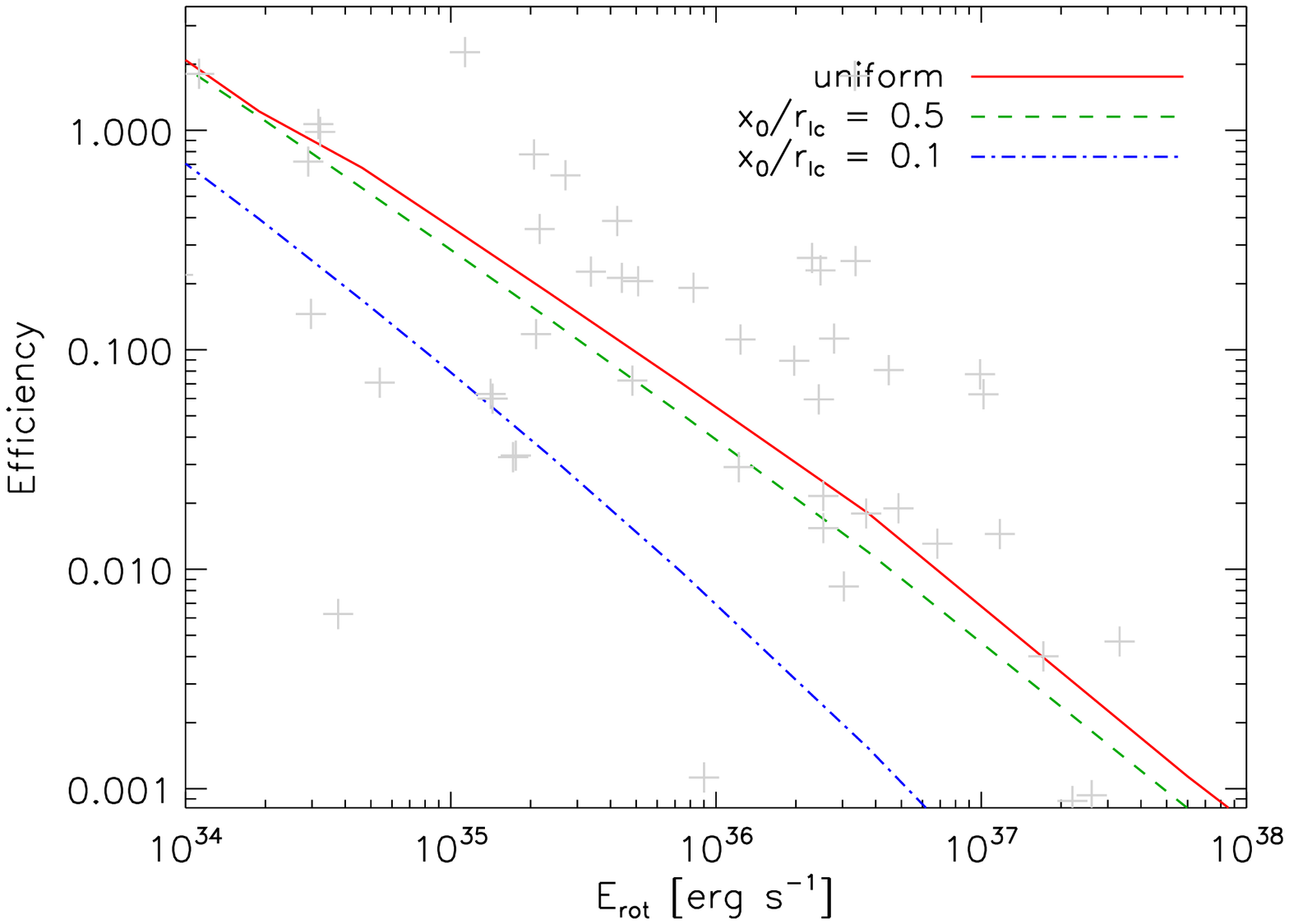}
\caption{$\gamma$-ray luminosity and efficiency for our models Y00, changing only the period, and for different effective particle distributions. Gray points indicates the observed young pulsars. See text for discussion.}
 \label{fig:efficiency}
\end{figure*}

In Fig.~\ref{fig:efficiency} we show the $\gamma$-ray 0.1--100 GeV luminosity, $L_\gamma$, and efficiency (defined as the ratio between $L_\gamma$ and the spin-down power $\dot{E}_{\rm rot}$) versus the rotational energy for our model Y00, as obtained by changing the period (i.e., the light cylinder) in our simulations. In our models we see a strong anti-correlation in these quantities. This is due to the fact that, keeping the same $E_\parallel$ but increasing the period (i.e., the light cylinder) the radiation losses are less effective. However, we recommend to take these results these with caution since, for instance, a dependence of $E_\parallel$ with the period would change this correlation. A change in $N_0$ would also re-scale luminosity and efficiency. Furthermore, the explored phase space parameters of the model does not always guarantee that the efficiency is less than one (see the lowest spin-down models at the top-left part of the diagram); although we see a general consistency and even a relatively good comparison with data --despite the significant uncertainty introduced by, for instance, distance estimates in the real pulsars.

Finally, we remark that the presented results are generally valid for any model relying on the synchro-curvature spectra. As a consequence, since we have not used specific characteristics of the OG, the possible constraints on the position of the emitting region can be applied to any kind of gap.

\section{Conclusions}\label{sec:conclusions}

In this work, we have calculated the synchro-curvature radiation for a variety of model parameters, with an analytical treatment which takes into account the uncertainties intrinsic to the assumptions of the model, as discussed in Paper~I.

First, our calculations of the optical depth for the pair production show how the X-ray flux and the position of the gap are important. The former depends on the size of the emitting region, which, in the thick outer gap model, is usually assumed to be limited to the bombarded polar cap. However, we recall that for young pulsars, the residual heat can make the entire surface hot, providing different temperatures compared to the bombardment estimate. The position of the gap depends on the radius of the light cylinder (the typical length scale of the problem), the global magnetospheric configuration, and the electrodynamics of the gap, which is quite uncertain.

Second, we dealt with expected spectra, motivated by the fact that the wealth of {\em Fermi}-LAT data \citep{2fpc} allows us to test phenomenological and physical gap models. In the literature, the modeling of $\gamma$-ray spectra has not been deeply explored. The most popular mechanism is thought to be the synchro-curvature radiation (often well-approximated to the purely curvature radiation). For a given set of magnetic field, curvature radius, electric field, then the emitted radiation is described by a cut-off power law, with a fixed low-energy slope of the energy spectral distribution $dP_{sc}/dE \sim E^{0.25}$. However, the observed {\em Fermi}-LAT pulsars show, on average, lower values of the power law index, between -1 and 0.25. To our knowledge, this common feature has not been explained yet, since spectra were computed for a single-value set of parameters, and considering only the steady-state, in which $\Gamma$ has reached the equilibrium value dictated by the balance between the electrical acceleration and the radiative losses, dominated by curvature. 

We have shown here that the qualitative features of the observed spectra can be reproduced by fully (using the full synchro-curvature losses) solving the particle dynamics, and considering the superposition of the radiation emitted at the different positions along the gap. In particular, if the viewing geometry, the position of the gap, and the beaming of radiation are such that the initial, high-$\xi$ part of the particle's trajectory, whose emission approximated by purely synchrotron radiation, weigh in more than the outer parts, we obtain softer low-energy slopes.

In order to investigate this feature in deeper detail and to gather generic constraints on the model parameters, we shall next use the wealth of {\it Fermi}-LAT data. In a subsequent paper, we shall fit models to observed spectra of individual sources. The spectral fits can give important indications on the radius of curvature and on the particle dynamics. The peak energy is an indicator of the magnitude of the electric field and the curvature radius. A systematic spectral fit could also be able to confirm synchro-curvature radiation as the dominant radiative mechanism in most pulsars. In this sense, there are two main features to be considered.

First, the low-energy slope and the broadness of the peak can constrain the effective distribution of particles. The flat low-energy slope seen in many pulsars is compatible with having more particles effectively emitting  in the inner part of the gap. Second, the sub-exponential dependence in energy after the energy peak, claimed for the phase-average, good-quality spectra of several sources, on the other hand, cannot be reproduced by a purely synchro-curvature spectrum, which has exponential dependence on $E$ after the peak. However, a possible explanation for at least some of the pulsars is that this is the result of the average over the phase of the emitted spectra, which parameters (like the peak energy and the slope index) can vary by factors of a few between thus providing a slower decay (see, e.g., \citealt{abdo10,abdo12}). On the other hand, if the sub-exponential cutoff is systematically seen and confirmed also for the phase-resolved spectra, and/or a power-law fits data above several tens of GeV \citep{aliu08,leung14}, it could point to other possible high-energy radiation mechanism: inverse Compton scattering against photons coming from the surface and/or other parts of the magnetosphere, or against the same $X$-ray synchro-curvature photons. The need for such component remains to be assessed in future works.

\section*{Acknowledgements}

This research was supported by the grant AYA2012-39303 and SGR2014-1073 (DV, DFT) and the Project Formosa TW2010005, for bilateral research between Taiwan and Spain (KH, DFT). KH is partly supported by the Formosa Program between National Science Council in Taiwan and Consejo Superior de Investigaciones Cientificas in Spain administered through grant number NSC100-2923-M-007-001-MY3. 
The research leading to these results has also received funding from the
European Research Council under the European Union's Seventh Framework
Programme (FP/2007-2013) under ERC grant agreement 306614 (MEP). MEP
also acknowledges support from the Young Investigator Programme of the Villum Foundation.
We are grateful to the referee for a careful reading and useful comments.

\bibliography{og}

\end{document}